\documentstyle[prb,aps,epsfig,multicol,amssymb,amsfonts]{revtex}
\tighten
\renewcommand{\narrowtext} 
{\begin{multicols}{2}\global\columnwidth20.5pc} 
\renewcommand{\widetext}
{\end{multicols}\global\columnwidth42.5pc} 
\newcommand{\be}{\begin{equation}}
\newcommand{\ee}{\end{equation}}
\newcommand{\bea}{\begin{eqnarray}}
\newcommand{\eea}{\end{eqnarray}}

\begin{document} 
\draft 
\title{The conductance of molecular wires and DFT based transport calculations} 
\author{F.~Evers, F.~Weigend, and M.~Koentopp} 
\address{Institut
f\"ur Nanotechnologie, Forschungszentrum Karlsruhe, 76021 Karlsruhe,
Germany}
\date{\today}
\maketitle
\begin{abstract}
The experimental value for the zero bias conductance of organic
molecules coupled by thiol-groups to gold electrodes tends to be much
smaller than the theoretical result based on density functional
theory (DFT) calculations, often by orders of magnitude.
To address this puzzle we have analyzed the regime within which
the approximations made in these calculations are valid. 
Our results suggest that a standard step in DFT based transport
calculations, namely approximating the exchange-correlation potential
in quasistatic nonequilibrium by its standard equilibrium expression, 
is not justified at weak coupling. We propose, that the breakdown of
this approximation is the most important source for overestimating
the width of the experimentally observed conductance peak and
therefore also of the zero bias conductance. We present a numerical
study on the conductance of the organic molecule that has recently
been studied experimentally by Reichert {\it et. al.}
\cite{reichert02} that fully agrees with  this conclusion.
\end{abstract}

\pacs{PACS numbers: } 
\narrowtext
\section{Introduction}
\begin{figure}[btp]
\begin{center}\leavevmode
\includegraphics[width=0.9\linewidth]{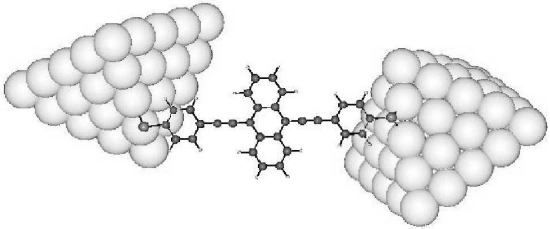}
\vspace*{-1cm}
\caption{ Schematic representation for a conductance measurement of the
molecule (9,10-Bis((2'-para-mercaptophenyl)\- -ethinyl)-anthracene):
between gold contacts [1]. }
\label{f1}\end{center}\end{figure}

Recently, several conductance measurements of single organic
molecules have been reported \cite{reichert02,reed97,cui01,park02,liang02}.
Fig. \ref{f1} illustrates a typical setup with one of the 
molecules used in experiments\cite{reichert02}.
\vspace*{-7.24cm}
\hspace*{1.75cm}
In part the tremendous attention that 
this field ``molecular electronics'' has received
in recent years is due to its possible
technological impact. Indeed, a molecular storage device
has 
\vspace*{5.4cm}
already been built \cite{roth00} and also molecular switches have 
been realized \cite{chen02}. It is clear that for technological applications
being able to systematically model and understand
the charge transport properties of the combined system of molecule
and contacts is of crucial importance.

Since the conductance is very
sensitive to details of spectral and orbital properties of
molecules and their wavefunctions, ab initio methods like density
functional theory (DFT) are an indispensable tool not only for
structure calculations but also for transport theory. 
Nowadays, DFT calculations including several thousand electrons are
possible, which allows the treatment of large molecules and also
to include parts of the contacts for more realistic calculations.
Despite these enormous capacities, a quantitative description
of transport for weakly coupled molecules with a conductance
well below the conductance quantum, $g{\ll}1$,
has not been achieved. In fact, experimental
and theoretical values for the zero bias conductance of organic
molecules, e.~g. benzene, often differ by 1-3 orders of
magnitude \cite{pantelides02,emberly01}.

In the present work we consider the organic molecule depicted 
in Fig. 1, which has been subject to experimental \cite{reichert02}
and DFT based theoretical investigations  \cite{heurich02}, 
before. 
In this case the experimental value for the zero bias conductance
is smaller than the theoretical one by a factor $\sim 20$.
Three different reasons that possibly could lead to this discrepancy
should be mentioned and discussed\cite{ghosh03}. 

First, Fig. 1 shows an idealized situation and actual microscopic
conditions realized in the experiment are not known.
However, we have chosen the particular case, Fig. 1, because
the experimental results are well reproducable \cite{reichert03}.
If important variations in the atomic structure of the contact
would exist, one should expect strong fluctuations in the conductivity
from sample to sample. Indeed fluctuations are present,
see Fig. \ref{f4} below, but their magnitude is relatively small
so that strong structural variations are unlikely to occur.
Small variations in structure, on the other hand, tend to have little
impact on the theoretical conductance, only. Our calculations
show, that the transmission is not very sensitive to changes in the bond angles or
bond lengths as long as the change is within reasonable limits.
Therefore, deviations from the assumed atomic arrangement
depicted in Fig. 1 from its experimental realization do not seem to
offer a plausible explanation of the large discrepancies observed.

Second, in theoretical calculations approximations have to be done
when the molecule is coupled to the leads. Artefacts can be excluded
only, when parts of the leads are included into the calculation
("extended molecule"), so that the following hierarchy of inequalities is met:    
\begin{equation}
  \label{ineq}
    \delta_{\rm eM} <  \gamma_{\rm eM}  \ll \gamma_{\rm M} 
\end{equation}
($\delta_{\rm eM}$: level spacing of extended molecule,
 $\gamma_{\rm M}$ ($\gamma_{\rm eM}$) level broadening of bare
 (extended) molecule when coupled to the leads).
Indeed, these conditions have not been met in previous calculations
\cite{heurich02}, where $\delta_{\rm eM}\approx \gamma_{\rm M}$.
In order to improve upon this result, we have performed
conductance calculations for the
extended molecule, Fig. 1, where $\delta_{\rm eM}/\gamma_{\rm M}
\approx 0.1$ and $\delta_{\rm eM}/\delta_{\rm M}\approx 0.01$
so that the hierarchy of inequalities is satisfied and artefacts from modeling of the leads can be 
excluded. We find that qualitative agreement of our data with the experiment is improved 
significantly as compared to the earlier calculation
\cite{heurich02}. However,
the quantitative disagreement is not decreased but rather increased by another factor of 10!  
This shows that insufficient modeling can be ruled out as a source
for the discrepancy. 

After we have discarded two obvious possibilities for
explaining the observed discrepancy, we have to resort to
more fundamental considerations. In this paper, we advocate
a third possibility, namely that the standard implementation
of DFT based transport calculations relies crucially on
assumptions that are not justified in the limit of
``weak coupling'' where the spatial structure
of molecular wavefunctions is strongly inhomogenous
and transport is dominated by single resonances. \cite{fn1} 

We give a outline of our paper and a brief account of key results. 
In section \ref{s2} we discuss 
the derivation of the basic equation of 
DFT-transport, Eq. \ref{e9}, and show that it has a wide
range of applicability. This is true provided that the DFT-calculations
are performed with an exchange-correlation potential appropriate
for the nonequilibrium situation. However, since this potential
is not known in general, in practice an ``equilibrium exchange
correlation approximation'' (EXCA) is made in which    
the standard equilibrium potential is used. Apriori, this
approximation  is uncontrolled. In section \ref{ss2.2} we
introduce an alternative -- but equivalent -- formulation
of transport, the {\it Kubo formalism}. It enables us to
give analytic arguments for the applicability of the EXCA. 
Our results suggest, that the EXCA can be trusted only
in the case of non-resonant transport, when the molecular
level broadening is strong and individual molecular levels
strongly overlap. A typical example for this case is a
chain of gold atoms. By contrast, if the transmission
exhibits resonances, transport is determined by individual
molecular orbitals and their broadening. In this case,
the corrections to the EXCA become significant and in particular
the level broadening and therefore the zero bias conductance
may be severely overestimated.
The molecule in Fig. 1 is a representative for these systems. 

In section \ref{s3}, we present numerical transport
calculations  for a gold chain and the molecule,
depicted in Fig. 1, that we have already alluded to above.
In view of our theoretical analysis we propose that it is
the breakdown of the EXCA in the weakly coupled limit that
is the cause of the large discrepancy between theoretical
and experimental molecular conductance. A discussion of our
findings will be given in section \ref{s4}

\section{Transport Formalisms}
\label{s2}
\subsection{Landauer-Buttiker-Formalism for Interacting Electrons}

Meir and Wingreen have derived a general expression
for the current flowing through a region of space, where the
charge carriers can interact -- like a quantum dot or a molecule\cite{meir92}.
Specifically, the Hamiltonian has the structure
\begin{eqnarray}
  \label{e0}
H = H_{0}(\{{\bf d}^\dagger_{m}\};\{{\bf d}_{m}\}) +
    \sum_{\alpha=l,r} \epsilon_{\alpha} {\bf c}^\dagger_{\alpha} {\bf
    c}_{\alpha} 
    \nonumber \\
+ \sum_{m \alpha=lr} \! (t_{\alpha m} {\bf c}^\dagger_{\alpha}
    {\bf d}_m + {\rm H.c.}).
\end{eqnarray}
The first term describes the bare molecule which in general may
include the electron-electron interaction. Its detailed structure
will be of no importance in what follows. 
The molecule makes contact to two leads denoted
left and right.
The bare leads are assumed to be noninteracting and
described by the second term in Eq. (\ref{e0}).
The third term represents the contact.

The Meir-Wingreen formula 
connects the retarded (advanced) Green's functions \boldmath $G$ ($G^\dagger$)
\unboldmath and the lesser function \boldmath $G^<$ \unboldmath
of the full many-body problem (including the leads) with the {\it dc}-current
\begin{equation}
  I {=} \int \!\! dE \ 
  {\rm tr} (f_{\rm L} \mbox{\boldmath $\Gamma_{\rm L}$\unboldmath} {-}
  f_{\rm R}
  \mbox{\boldmath $\Gamma_{\rm R}$\unboldmath})
  (\mbox{\boldmath $G-G^\dagger$\unboldmath})
  +
  {\rm tr} (\mbox{\boldmath $\Gamma_{\rm L}$\unboldmath} {-}
  \mbox{\boldmath $\Gamma_{\rm R}$\unboldmath})
  \mbox{\boldmath $G^<$\unboldmath}
  \label{e1}
\end{equation}
\noindent	   
where $f_{\rm L,R}=f(E-\mu_{\rm R,L})$ denotes the Fermi distribution
functions for the leads at chemical potentials $\mu_{\rm R,L}$ and
\boldmath  $\Gamma_{\rm L,R}$
is the imaginary part of the self energy that describes the coupling of the molecule to the
external leads. (For the most part the dependence on energy
\unboldmath$E $ \boldmath will be suppressed in our notation.) 
For {\it noninteracting} particles, in terms of these self energies we have 
 \boldmath${G}^{-1}$\unboldmath$=E$\boldmath$1-$\unboldmath 
      \boldmath$H_0$\unboldmath$-$\boldmath$\Sigma_{\rm L}$\unboldmath$-$\boldmath$\Sigma_{\rm R}$,
so that 
 $\Gamma_{\rm L,R}=\mbox{\unboldmath{$\imath$}}(\Sigma_{\rm
L,R}-\Sigma_{\rm L,R}^{\dagger})$     
and
\boldmath
\begin{equation}
  G^< = \imath G \ ( \ \mbox{\unboldmath $f_{\rm L}$ \boldmath}
  \Gamma_{\rm L}
  +  \mbox{\unboldmath $f_{\rm R}$ \boldmath} 
  \Gamma_{\rm R} \ ) \ G^\dagger.
\label{e2}
  \end{equation}
\unboldmath
\noindent	   
In the appendix, we demonstrate that Eq. (\ref{e2}) is just the statement,
that the density matrix can be constructed from left and right going
scattering states $\psi_{l,r}$ ($x=({\bf x},t)$),
\begin{equation}
\boldmath{G^<}\unboldmath(x,x') = 
\sum_{l,r} \ f_l \  \psi_l(x) \ \psi^*_l(x') +  
           f_r \ \psi_r(x) \ \psi^*_r(x').
\label{e3}
\end{equation}
\noindent	   
From (\ref{e1}) together with
(\ref{e2}) we find for the transmission in the limit of linear
response and zero temperature
\begin{equation}
T(E)={\rm tr}\ \mbox{\boldmath$\Gamma_{\rm L} G$\unboldmath}
\mbox{\boldmath$\Gamma_{\rm R} G^\dagger$\unboldmath}.
\label{e4}
\end{equation}
\noindent
Eq. (\ref{e4}) is an incarnation of the familiar Landauer-Buttiker
formula \cite{imry} and valid for noninteracting electrons.

In this section, we argue that expression (\ref{e4}) continues
to hold also for a much larger class of interacting problems.
This is because according to the Runge-Gross theorem of time
dependent DFT (TDDFT) the time
evolution of any many-particle Hamiltonian can be calculated by
solving a single particle problem in an appropriate
effective potential\cite{onida02}. 
Since this point is of importance to us we elaborate on it
in the following.

The effective time dependent single particle problem we should solve
is of the form
\begin{equation}
  -\imath \frac{d}{dt} \theta_n(x) = (-\frac{1}{2m} \nabla^2 + V_s(x))
   \theta_n(x)
   \label{e5}
  \end{equation}
where
\begin{equation}
  V_s(t) = V_{\rm i} + V_{\rm H}[n] + V_{\rm
  XC}[n] + V_{\rm ex}(t)
  \label{e6}
  \end{equation}
with $V_{\rm i}$ denoting the ion-core potential, $V_{\rm H}$ the
Hartree-interaction, $V_{\rm XC}$ the exchange-correlation potential
and $V_{\rm ex}$ the external probing field. (Notation suppresses the
spatial index ${\bf x}$.) In addition to the
explicit time dependence of the potential $V_s$ imposed by
$V_{\rm ex}$ an implicit dependence exists because $V_s$ is a
functional (in general nonlinear and nonlocal in time and space)
of the density $n({\bf x},t)$. Let us assume that Eq. (\ref{e5})
describes the molecule together with the leads and an
external perturbation $V_{\rm ex}$ that is switched on
at $t=t_0$ and time independent thereafter: 
$$
V_{\rm ex}(x) = V_{\rm ex}({\bf x}) \Theta(t-t_0)
$$
At times prior to $t_0$ the system is assumed to be in thermal equilibrium
($T=0$), so that the reservoirs are characterized by occupation
numbers $f_n$. 
Then we have $n(x) = {\cal G}^<_V(x,x)$, with
\begin{equation}
  \boldmath{\cal G}^<_V \unboldmath (x,x') = \sum_{n} f_n \ \theta_n(x) \theta^*_n(x')
  \label{e7}
    \end{equation}
where $\theta_{n}({\bf x},t)$ evolves according to Eq. (\ref{e5}).
Moreover, we specialize to the case,
where $V_{\rm ex}({\bf x})$ generates a monotonous electrical potential drop
from $V_{\rm ex}=V$ in the asymptotic region of the left lead to
$V_{\rm ex}=0$ in the right lead. In response to the potential
drop a current is being generated. After an initial period 
exhibiting transient behavior, there will be a parametrically wide
time interval in which the current and the density are
quasistationary. This is precisely the situation for which
also Eq. (\ref{e1}) has been derived\cite{meir92}.
It is only at even much longer times, that the
electrochemical potential becomes homogenous again and the current
stops to flow. Formally speaking, we perform the order of limits
in which the size of the reservoirs is send to infinity first and
$t_0\rightarrow -\infty$, thereafter. We mention that,
since we are interested in the long-time-behavior only,
details of how the external potential is switched on,
step-like or adiabatic, are unimportant.
The corresponding memory is erased inside the reservoirs.

Since Eq. (\ref{e7})
describes the exact evolution of the time dependent density
we can also calculate the (longitudinal) current density $j=dn/dt$
and hence find an {\it exact} expression for the conductance.
Indeed, the same reasoning that is used
for noninteracting electrons to relate transport to a
scattering problem can be employed for the present effective single
particle problem as well. Consequently, a zero temperature description
of the quasistationary region in terms of scattering states
should exist with ${\cal G}^<_V$ 
taking a form similar to Eqs. (\ref{e2}) and (\ref{e3})
\begin{equation}
{\cal G}_V^<(x,x') = 
\sum_{l,r} \ f_l \  \theta_{l}(x) \ \theta^{*}_{l}(x') +  
           f_r \ \theta_r(x) \ \theta^{*}_r(x').
\label{e8}
\end{equation}
\noindent	   
It involves the Kohn-Sham orbitals $\theta_{r,l}$ representing the
scattering states of the quasistationary nonequilibrium situation
and their zero temperature occupation numbers $f_{l,r}$ imposed by 
the left and right reservoirs.
Since also the derivation of Eq. (\ref{e2}) can be repeated for an effective
single particle problem, a relation analogous to (\ref{e1})
also holds for ${\cal G}^<$ and the corresponding retarded Green's
function ${\cal G}$.
Together with the previous statement (\ref{e8})
this implies that the transmission is given by
\begin{equation}
T = {\rm tr} \ \mbox{\boldmath $\Gamma_{\rm L}{\cal G}_V$\unboldmath}
   \mbox{\boldmath $\Gamma_{\rm R} {\cal G}^\dagger_V$\unboldmath}     
     \label{e9}
\end{equation}
with \boldmath${\cal G}^{-1}_V$\unboldmath$(E)=$
      \boldmath${\cal G}^{-1}_{0V}$\unboldmath$-$\boldmath$\Sigma_{\rm
	L}$ \unboldmath$-$\boldmath$\Sigma_{\rm R}$\unboldmath,
and the resolvent matrix
\begin{equation}
\mbox{\boldmath${\cal G}_{0V}$\unboldmath}({\bf x},{\bf x'},E) = \sum_{n}
\ \frac{\phi_{n}
  ({\bf x})\phi^*_n({\bf x'})}{E-\epsilon_n + \imath 0}
\label{e10}
\end{equation}
where the sum is over KS-energies $\epsilon_n$ and orbitals
$\phi_n({\bf x})$ calculated for the uncoupled molecule.

Eq. (\ref{e9}) constitutes the main result of this section. Similar
arguments can also be found in a recent communication
by Stefanucci and Almbladh\cite{stefanucci03}. 

Several aspects of our finding should no
pass by without further comment.
a) Eq. (\ref{e9})  constitutes the generalization
of the Landauer formula to interacting
electron systems. In the special case, where
\boldmath$\Gamma_{\rm L}$ \unboldmath and
\boldmath $\Gamma_{\rm R}$ \unboldmath
differ by a constant factor only, a derivation
has already been given by Meir and Wingreen before\cite{meir92}.
We emphasize, however, that this condition is extremely restrictive.
It implies that every atom of the molecule is coupled in precisely
the same way to the left lead as it is coupled to the right lead.
Given the fact that physical couplings decay with increasing
spatial distance, the ``condition of proportional couplings''
is violated for every realistic system with a finite extent.

b) As before the self energies
\boldmath$\Sigma_{\rm L,R} $ \unboldmath
account for the coupling of the molecule to the leads. 
It is very well known that 
they can incorporate sophisticated many body effects,
like i.e. the Abrikosov-Suhl resonance
if Kondo-physics is present\cite{mahan}. However, one
can restructure the problem and define an ``extended molecule''
that also comprises parts of the contacts, see Fig. 1.
The new contact surface ${\cal S}_{\rm eM}$ can be arranged sufficiently far away
from the physical contact, so that the new self energy
depends on the type of lead only, but is totally independent
of which molecule is used. In fact, if a separation of energy scales
exists such that the level spacing of
the extended molecule, $\delta_{\rm eM}$,  is much smaller than 
the broadening, $\gamma_{\rm M}$,  of the bare molecular levels
upon coupling to the leads the microscopic information carried by 
\boldmath$\Sigma_{\rm L,R} $ \unboldmath becomes irrelevant.
With any choice of $\gamma_{\rm eM}$ in accord with Eq. (\ref{ineq})
the simple replacement
\boldmath$\Sigma_{\rm L,R}$\unboldmath({\bf x,x'})
$=\imath\gamma_{\rm eM} \ \delta_{\bf x x'}$ \unboldmath
with ${\bf x},{\bf x'}$ situated on ${\cal S}_{\rm eM}$
is justified. This freedom merely reflects the
experimentalists choice to attach leads at convenience
from any sort of shape or material as long as the voltage drop
is near the real molecule. 
In this limit, all the detailed information about transport
properties of the molecule including correlation effects
is carried by the resolvent matrix (\ref{e9}) and
inherited from the exchange correlation potential. 

c) Eq. \ref{e9} has been derived under the condition of vanishing
   temperature and frequency and in the regime of linear response.
   Under these restrictions, scattering is energy conserving and
   therefore an effective single
   particle scattering formulation of transport can exist. Upon releasing the
   constraints the incoming particle can exchange energy with the
   molecule. Qualitatively new phenomena can occur and in general
   the scattering problem will become much more complicated and may
   not be understood in terms of a simple single particle picture. For
   example memory effects appear, because here the incoming electron sees
   the molecule in the state it has been left in after interaction
   with the previous electrons.

d) Eq. (\ref{e10}) suggests that near equilibrium
(i.e. for the linear response) a variational principle may exist
that allows for the calculation of the voltage at a given current
or vice versa. The idea is to introduce a density operator that
maximizes the entropy under the constraint that the current be finite.
Work along this line has been done by Kosov\cite{kosov03}, recently.

\subsection{Kubo Formalism of Linear Response}
\label{ss2.2}

The  practical usefulness of Eq. (\ref{e9}) is limited since not
much is known about the exchange correlation potential $V_{\rm XC}(V)$ that
defines the Kohn-Sham problem in quasistatic nonequilibrium. 
Throughout all previous works using the DFT approach to transport
it has been universally assumed that $V_{\rm XC}(V)$ may be
approximated by its equilibrium form $V_{\rm XC}(0)$ 
used in standard DFT-calculations (EXCA).
Let us discuss now, under which conditions this
approximation may be expected to hold.

Instead of solving a scattering problem one can also find
the conductance from the alternative, but equivalent,
Kubo formalism. 
The advantage of this starting point for our purposes is that
analytical statements about the excitation frequencies are available
which contain information about the corrections to the EXCA.

We observe, that the current is related to the
{\it dynamical} polarization
\begin{equation}
   {\bf I} = d{\bf P}/dt.
   \label{e11}
  \end{equation}
As a consequence the
linear current response to a homogenous external electric field
is governed by the dynamical polarizability tensor of the entire system including
molecule and leads
\begin{equation}
  {\bf I} = -\imath \omega  \mbox{\boldmath
    $\alpha$\unboldmath}(\omega)\ {\bf E_{\rm ex}}.
  \end{equation}
which is closely related to the full density susceptibility 
\begin{equation}
  \alpha_{ij}(\omega) = \int d{\bf x} \ d{\bf x'} \ x_i \ x'_j\
  \chi({\bf x},{\bf x'}, \omega)
  \label{e12}
    \end{equation}
with the corresponding Lehmann representation
\begin{equation}
  \alpha_{ij}(\omega) = \sum_{\mu=0}
  \frac{\langle 0|\hat x_i|\mu\rangle \langle \mu| \hat x_j|0\rangle}
       {\omega - E_{\rm \mu 0}+\imath0}.
       \label{e13}
       \end{equation}
\noindent
The sum is over all many body states with energies $E_\mu$
starting from the ground state $\mu{=}0$ and the abbreviation
$E_{\mu 0}{=}E_\mu{-}E_0$ has been used. In the case of strongly
resonant transport the sum is dominated by the contribution
of a few poles and the off-resonant current results from  the
finite lifetime of the corresponding excitations which is encoded
in the dipole matrix elements. Clearly, this
is the regime in which molecules attached
with thio-groups find themselves.

The full density response contains a
piece that describes {\it dynamical screening}.
In terms of the bare KS-response
$\chi_{\rm KS}({\bf x},{\bf x'},\omega)$
the relation
\begin{equation}
  \chi^{-1}(\omega) = \chi^{-1}_{\rm KS}(\omega) -
  v_{\rm Q} - f_{\rm XC}(\omega)
  \label{e14}
  \end{equation}
holds true, where the exchange-correlation kernel
$f_{\rm XC}(x,x') =dV_{\rm XC}(x)/dn(x')$ has been introduced
and $v_{\rm Q}=|{\bf x}-{\bf x'}|^{-1}$. In fact, Eq. (\ref{e14})
can be thought of as a definition of $V_{\rm XC}$.
The crucial point is that the additional terms in (\ref{e14})
shift the excitation energies $\Omega_{\mu}$ away
from their bare KS-values
$\omega^{\rm KS}_{ij}=\epsilon_{i}-\epsilon_{j}$
(we introduce a multi-index $\mu=(i,j)$)
\begin{equation}
  \Omega_{\rm \mu} \approx \omega^{\rm KS}_{\rm \mu} + \langle \mu | \
  v_{\rm Q} |\mu\rangle + \langle \mu | f_{\rm XC} (\omega^{\rm KS}_\mu) | \mu \rangle
\end{equation}
where $|\mu\rangle=|i,j\rangle$ denotes a single particle transition
from the occupied KS-orbit $j$ to the vacant orbit $i$.
When $f_{\rm XC}$ is decomposed into its exchange
contribution and the correlation part $f_{\rm C}$ further
analytical progress is possible \cite{gonze99} and one finds
\begin{eqnarray}
  \Omega_{\rm \mu} &\approx& \omega^{\rm KS}_{\rm \mu} +
  \langle j | \hat v^{\rm HF}_{\rm x} - v^{\rm KS}_{\rm x}| j \rangle
  - \langle i | \hat v^{\rm HF}_{\rm x} - v^{\rm KS}_{\rm x}| i \rangle \nonumber\\
  && + \langle \mu | \ v_{\rm Q} |\mu\rangle 
  - \langle ii  | \ v_{\rm Q} | jj \rangle
  + \langle \mu | f_{\rm C} | \mu \rangle.
  \label{e16}
\end{eqnarray}
($\hat v^{\rm HF}_{\rm x}$: exchange part of the bare Hartree-Fock
(HF) operator,
$v^{\rm KS}_{\rm x}$: exchange potential in the KS-equation).
Eq. (\ref{e16}) contains the plausible statement that the actual
excitation energies move away from their KS-values
closer to their Hartree-Fock estimates when dynamical
screening is taken into account \cite{gonze99}. 
Note, that a corresponding change must be expected to happen also in
the dipole matrix elements of Eq. (\ref{e13}) which can be important
since they control the lifetime of the excitations.
  
Equipped with this information we can proceed now and discuss the validity
of the EXCA underlying previous DFT-transport calculations.
Two limiting cases should be distinguished:

i) individual resonances strongly overlap, $\gamma_{\rm M}\gg \delta_{\rm M}$.  
   This is the limit of strong coupling, $g\sim 1$,
   where the wavefunctions $|i\rangle$ are extended and 
   the deviations from KS- and HF-excitation energies in
   Eq. (\ref{e16}) are small. In fact, for structureless,
   plane-wave-like states they vanish. In this case the
   standard approximation is justified. An example for
   a system in this class is a linear chain made from
   Au-atoms. 

ii) individual resonances are fully developed, $\gamma_{\rm M}\ll \delta_{\rm M}$.  
   Here, the wavefunctions show pronounced localized features and
   in general the corrections to KS-excitation energies and matrix
   elements will not be small. While qualitatively correct results
   may still be found in many cases, a systematic quantitative
   analysis of the resonant structures in $T(E)$ based on the EXCA
   is not possible, in general. A typical representative
   of weakly coupled systems is given in Fig.~\ref{f1}. The
   contact of the extended $\pi$-system to the leads is interrupted by
   an $S$-atom and therefore the molecular states are partly
   localized.

\section{Numerical Calculations}
\label{s3}

The discussion presented in the previous section suggests 
that the experimental conductance of Au-chains is well described by the
standard DFT-approach whereas the conductance of organic molecules
coupled by thio-groups is only poorly captured. 
In the following section we present transport
calculations making use of Eq. (\ref{e10})
together with the EXCA in order to
corroborate this result.

\subsection{Method}

Our method is similar to approaches described in
\cite{datta01,brandbyge02,Xue02}, however it has the advantage
that using the program package TURBOMOLE we can include a considerably larger
number of contact gold atoms. For the molecules of interest to us,
this number (110) is sufficient so that the Fermi energy of the
extended molecule is very close to the bulk value
$E_{\rm F}\approx -5.1$eV, even without attaching additional leads.
 
Since the jest of the method has been outlined previously
\cite{datta01,brandbyge02,Xue02} we can limit ourselves to a brief
description. The transmission is given by Eq. (\ref{e9})
but with a resolvent matrix \boldmath${\cal G}_0$ \unboldmath that
has been obtained employing the EXCA: KS-orbitals and KS-energies
are taken from a standard DFT calculation \cite{turbomole,becke88}.
\boldmath
The self energies,
$\Sigma_{\rm L,R}$ can be expressed in terms of the
hopping matrix elements,
\mbox{\unboldmath$t_{\rm
L,R}(\mbox{\boldmath${\bf X}$}N,\mbox{\boldmath${\bf x}$}n)$}
which describe a hopping process of an electron in an orbital state
\mbox{\unboldmath$N$} of an atom at position ${\bf X}$
of the extended molecule to a state \mbox{\unboldmath$n$} of the
atom at a location ${\bf x}$ in the leads:
$\Sigma_{\rm L}=t_{\rm L} g_{\rm L} t^\dagger_{\rm L}$ and similarly for
$\Sigma_{\rm R}$. The hopping matrix elements we approximate by their
bulk values that we obtain from an independent DFT calculation for
a large gold cluster (146 atoms). Likewise, we replace the surface
Green's function $\langle {\bf x}\mbox{\unboldmath$n$}| g_{\rm L,R}|
{\bf x}\mbox{\unboldmath$^\prime$}\mbox{\unboldmath$n^\prime$}\rangle$ of the leads by a bulk
\unboldmath
one taken from the same calculation
\cite{tobepublished}. 

\vspace*{-0.5cm}
\subsection{Results}
\begin{figure}[btp]
\begin{center}\leavevmode
\includegraphics[width=0.79\linewidth, angle=270]{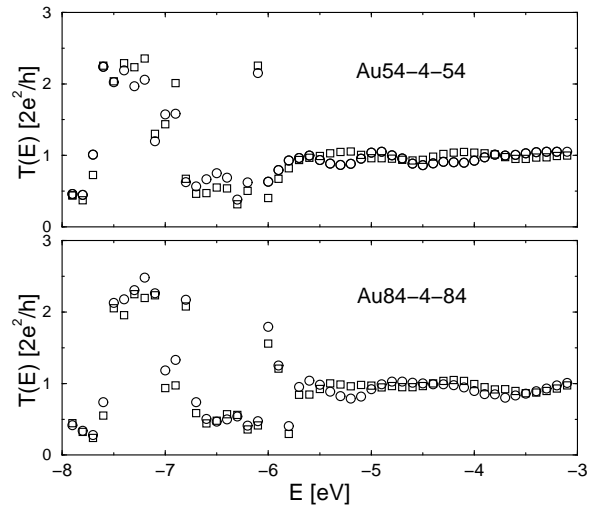}
\caption{Transmission of a four atom gold chain. Upper panel:
Contact was made of 54 gold atoms, see contacts in Fig.\ref{f1}.
Out of these 41 atoms ($\Box$, $5\times5$ and $4\times4$ bottoms layers)
have been coupled to external leads. For comparison, results with
29 coupling atoms are shown as well ($\circ$). Lower panel:
Same as upper panel with 84 gold atoms defining the contacts.
($E_{\rm F}\approx -5.1$) Traces are (nearly) independent of the
modeling parameters chosen, and they agree well with experiments and
previous calculations\cite{palacios02}} 
\label{f2}\end{center}\end{figure}

Fig. \ref{f2} shows
our results for the transmission of a linear chain of four
equidistantly placed Au-atoms ($d=2.67\mathring{A}$).
As a check we have performed  calculations 
with leads that have been modeled
by 54 (see Fig. 1) and 84 Au-atoms .
Moreover the number of ``surface'' Au-atoms from the extended molecule
that have been coupled to external leads has been varied:
the self energy has been calculated with 29 and 41 atoms
taken into account. It can be seen that the transmission is
(essentially) independent of these parameters, as it should. 
Our results are in very good agreement with experiments and
previous calculations \cite{palacios02}.

Before we present our results for the transmission function
the coupling of the organic molecule to the contacts
should be discussed. 

\vspace*{-1cm}
\begin{figure}[btp]
\begin{center}\leavevmode
\vspace*{-0.1cm}
\includegraphics[width=0.48\linewidth, angle=270]{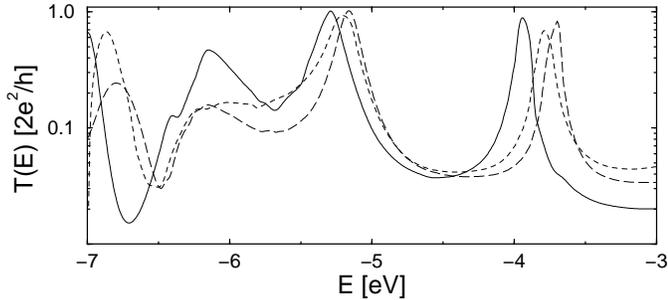}
\caption{
Transmission of the molecule in
Fig. \ref{f1} with sulfur atom coupling to one (solid),
two (dashed) and three (dotted) gold atoms.
($E_{\rm F}\approx -5.1$eV) Modifications in the atomic conact
structure lead to a slight shift of the conductance peak that
offer an explanation for the inhomogenous peak broadening observed
in experiments, see Fig. \ref{f4}.}
\label{f3}\end{center}\end{figure}
\vspace*{-0.5cm}
\noindent
In experiments a standard way to
facilitate this coupling is to introduce a 
sulfur atom that forms a very strong bond with
gold as well as with carbon atoms. 
Since the precise microscopic conditions of experiments are not
known, several possibilities for the sulfur-gold bonds have to
be considered. It is known, that
sulfur tends to bind to three gold atoms on a plane Au(111) surface,
i.~e. the hollow site is the most stable one\cite{sellers93}.
Binding to just one gold atom -- on top position -- corresponds to
a local minimum of the free energy and binding to two gold
atoms is unstable.
We find that the situation on rough surfaces exhibiting  edges,
is different. Near an edge, the sulfur finds its most stable position by binding
to two gold atoms (Fig. \ref{f1}) in agreement with an earlier study
on much smaller systems \cite{reimers02,weber02}.
The on top position remains a minimum albeit at
higher energy ($+0.7$eV) and the hollow position 
is unstable. Because in the break junction experiments of interest to us  the 
sulfur atom is likely to be exposed
to an irregular surface we consider all three cases.

\begin{figure}[btp]
\begin{center}
\vspace*{-0.75cm}
\includegraphics[width=0.55\linewidth, angle=270]{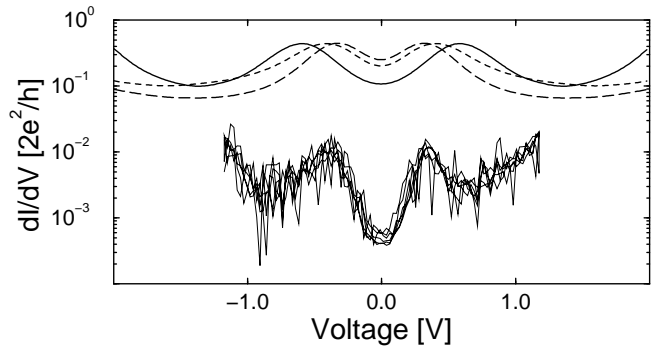}
\caption{Upper curves: dI/dV curve from data in Fig. \ref{f3}.
based on $I(V)=\int\! dE\,T(E)\,(f(E+V/2)-f(E-V/2))$ with $f(E)$
denoting the Fermi-function ($T=300$K).
Lower curves: experimental data from [1].
Different traces represent results from consecutive voltage sweeps.
The theoretical and experimetal traces  exhibt the same qualitative
features, but strong quantitative deviations in the conductance
magnitude exist. }
\label{f4}\end{center}\end{figure}

In Figs. \ref{f3} and \ref{f4}
we show the transmission and the I/V-characteristics
of the molecule of Fig. \ref{f1}. The traces correspond to the 
S-atom binding to one, two or three Au-atoms.
These micro-deformations induce a slight shift of
the transmission peak near $-5.2$eV by about $0.2$eV. 

Qualitatively, the observed features are in
accord with experimental findings of [1],
inset of Fig. \ref{f4}:
we find a conductance gap and the differential
conductance exhibits 
a maximum at about $0.4$V that
stems from a resonant molecular level - the remnant of the
HOMO of the bare molecule -
about $0.2$eV below the Fermi energy. The peak conductance and
the zero bias conductance can differ by an order of
magnitude and at larger voltages the conductance rapidly decreases
before there is another increase, again.
On the quantitative level serious discrepancies between our
calculations and experiments persist:
our value for the zero bias conductance $0.2 e^2/h$ exceeds the experimental one,
$\sim 10^{-3} e^2/h$, by more than two  orders of magnitude. Moreover, the
width of the experimental peak 
is strongly temperature dependent and can decrease 
\vspace*{-0.5cm}
\noindent
by an
order of magnitude with temperature decreasing from 300K to 30K \cite{reichert03}.
This strongly suggests that the experimental peak width at room temperature
is determined by inhomogeneous broadening rather than a
strong coupling to the leads. 

The inhomogenous broadening can be understood on the basis
of our results in Fig. \ref{f3}. 
We have seen above that the position of the HOMO is roughly
accounted for by our DFT procedure and it is mainly its damping
that is over\-estimated. Therefore, the shift of the peak position
upon changing the coupling of the molecule to the
leads is indicative of a microscopic
smearing mechanism: at sufficiently 
large temperatures the thermal average over the different types of 
S-Au-bonds leads to an effective broadening on the energy
scale $\approx 0.3$eV which indeed is in
accord with experimental observations.

\begin{figure}[btp]
\begin{center}\leavevmode
\includegraphics[width=0.5\linewidth]{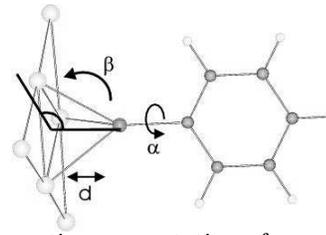}
\caption{Schematic representation of geometric degrees of freedom
defining the sulfur-gold-coupling. (center atom: sulfur, left: gold
surface, right: benzene-molecule)}
\label{f5}\end{center}\end{figure}

One might suspect that at least part of
the reason for the large discrepancy between theoretical and
experimental findings for the zero bias conductance is
that in real experiments the molecule is
exposed to various boundary 
conditions, e.~g. stress,
that prohibit the contact geometry to relax completely, which is
what we have assumed in our calculation.
Therefore, an important question is whether the conductance
can be strongly  affected by a slight change in microscopic
degrees of freedom, like the S-Au-bond length or the
bond angles defined in Fig. \ref{f5}.
 
\noindent
\begin{figure}[btp]
\begin{center}\leavevmode
\includegraphics[width=0.85\linewidth]{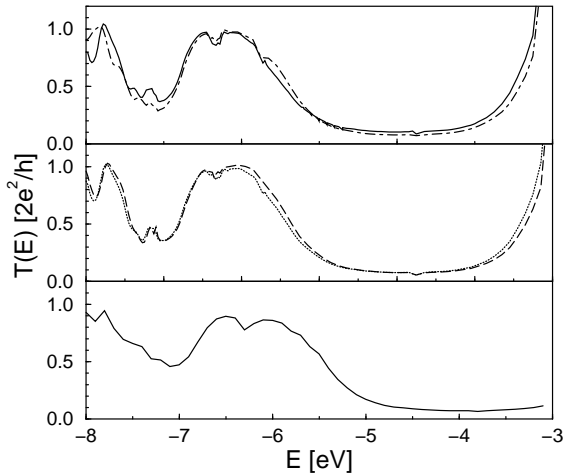}
\vspace*{0.6cm}
\caption{Transmission of benzene-1,4-dithiol for various parameters
$\alpha$, $d$ and $\beta$ as defined in Fig. \ref{f4}. Plot shows,
that the transmission is robust against small structural
modifications induced e.g. by strain.  
Upper panel: molecule, coupling to a Au(111) hollow site,
fully geometry optimized (solid) and same molecule
after rotation (dot-dashed, $\alpha=\pi/24$). Middle:
change in bond length (optimal value: $d=2.55${\AA}).
applied change $0.05${\AA} (dashed)
and $0.1${\AA} (long dashed). Bottom: change in angle $\beta$ by
$\pi/12$. ($E_{\rm F}\approx -5.1$eV)
}
\label{f6}\end{center}\end{figure}
\vspace*{-0.5cm}
Since our interest is in a qualitative question, we perform our
study using the simpler molecule benzene-1,4-dithiol.
Our findings for the transmission as a function of energy are
in excellent agreement with recent results by Xue and Ratner
\cite{Xue03} and Stokbro {\it et al.}\cite{stokbro03-2}.
In Fig. \ref{f6}, upper panel,  we display the
transmission of this molecule for the geometry optimized
case and also after a subsequent rotation of the molecule
about angle $\alpha=\pi/24$ defined in Fig. \ref{f4}.
The middle panel shows the impact of changing the sulfur-gold
bond length $d$ by 0.05{\AA} and 0.1{\AA}, the lower channel
exhibits the change upon changing the bond angle $\beta$ by $\pi/12$..
These manipulations have only a small effect on the transmission
confirming earlier findings \cite{Xue03,yaliraki99}.
The example shows that the DFT-conductance does not change by orders of
magnitude when changing details of the sulfur-gold bonds
within reasonable limits.

Now, we are facing the following situation:
On the one hand, the present calculations show 
that small variations in the nanostructure do not strongly
influence the transmission. On the other hand, experiments are
reproducible and do not show very strong fluctuations in the
zero bias conductance, which implies that large variations
in the atomic structure of molecule and contact do not exist.
Furthermore, the chemical bond between S and Au is
known to be very strong and stable -- this is why sulfur has been
chosen as the coupling element in the first place. 
Combining these three facts, we conclude that at present
there is little evidence that modifications in the atomistic
structure of the experimental contacts could exist,
which are not properly included in the theoretical modeling,
and therefore would hint to an explanation as to
why the theoretical conductance exceeds the experimentally
measured one so much.

\begin{figure}[btp]
\begin{center}\leavevmode
\includegraphics[width=1\linewidth]{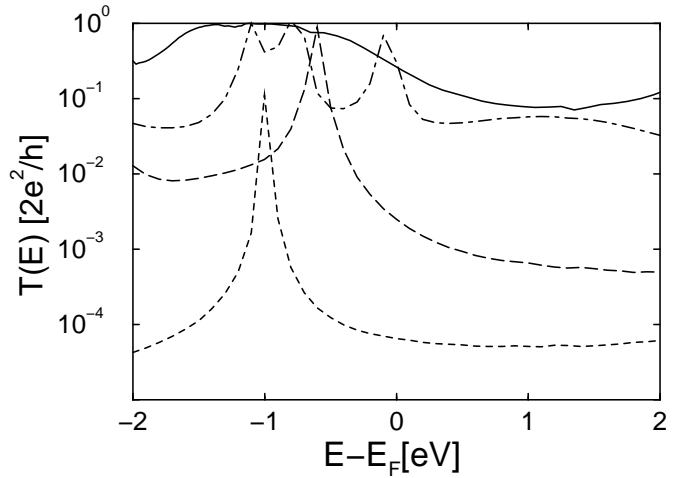}
\caption{Transmission of benzene-1,4-dithiol calculated
within DFT (solid Au54, as in Fig. \ref{f6}; dot-dashed Au14),
HF (dashed Au14) and extended H\"uckel (dotted Au14)
with 14 and 54 atom gold clusters simulating the electrodes.
DFT- and HF-calcuations show large differences which indicates
a breakdown of the EXCA explicated in section \ref{ss2.2}.
}
\label{f7}\end{center}\end{figure}
Instead, we advocate another resolution of the puzzle, namely that the
mechanism described in section \ref{s2} becomes active because
deviations from KS-states and HF-states are large.
To illustrate that the proposed mechanism may indeed induce
quantitative changes by orders of magnitude, we compare the
transmission calculated within the approximations DFT, HF
and extended H\"uckel. Fig. \ref{f7} exhibits the expected behavior:
the DFT-transmission peaks are shifted as compared to the Hartree-Fock
result and overlap much more strongly. In addition, the DFT-transmission
has a broad background contribution only weakly depending on energy.
It stems from electronic states that reside in the leads and
``leak'' into the molecule. The DFT-calculation overestimates
tunneling of these states as well. The net result is a difference
in the zero bias conductance by two orders of magnitude depending
on whether the DFT-approximation or the HF-approximation
is being used. Note, that because of the theoretical argument
given in section \ref{ss2.2} it is not entirely clear, apriori,
that it is necessarily DFT that gives the better approximation.

In order to reduce the numerical effort we
have restricted ourselves to small electrodes consisting of
14 Au-atoms each. By comparing the DFT-results for Au14 and Au54
depicted in Fig. \ref{f6} one can convince oneself, that
this simplification leaves unaffected our main conclusion.
\vspace*{-0.5cm}

\section{Conclusions}
\label{s4}

The numerical results presented in the preceeding section are in
full agreement with expectations based on the theoretical
analysis performed in section \ref{s2}:
while the conductance of a Au-chain agrees well 
with experimental results, 
large quantitative discrepancies
for organic molecule attached to gold electrodes via a
thiol-bond exist. Arguments have
been presented to the extent that these discrepancies
cannot plausibly be explained by insufficient modeling.

The standard DFT-approach can be exact only
in the case, where the ground state is represented by a
single Slater determinant which in turn implies that
HF is exact. For the molecule
this is not the case, however. Since an approximation
for the ground state based on the Slater-determinant of
KS-orbitals is a mixture of the wavefunction of the exact
ground state with its excitations, the character of the
approximate state tends to be too delocalized.  Consequently,
the molecule-lead coupling comes out too strong and the
level broadening is overestimated. We have given analytical
arguments in favor of this picture.

In order to improve upon the standard DFT-approach to transport
the exchange correlation potential should be replaced by the
appropriate potential for the quasistatic nonequilibrium in order
to include the corrections due to {\it dynamical} screening that are
ignored otherwise. Unfortunately, at present such a potential is not known.
However, our work suggests an alternative approach which is
based on the Kubo formula. It enables us to to calculate
the dynamical polarization of the molecule together with parts of
the leads. Thereof we can obtain the {\it dc}-current in the limit
of zero frequency. Work in this direction is under way.

Finally, let us emphasize that our results have implications for the 
calculation of nonequilibrium effects \cite{brandbyge02}
like the polarization of those electronic states that
do not carry the current. Two effects that work in opposite directions
occur. For a given current the voltage drop at the molecule
is underestimated by DFT-based calculations (since $g$ is too large)
and therefore so is the induced polarizing field.
On the other hand, the charge response of localized states to this field
is overestimated since they appear too metallic.
Consequently, interpreting  
DFT based transport calculations for nonequilibrium effects
is not straightforward and possible  asymmetries in the
I/V-characteristics could be blurred. \cite{stokbro03,romeike02}

\acknowledgements

We thank R. Ahlrichs, F. Furche,
A. Mildenberger, R. Narayanan, J. Reichert, H.~B. Weber and in
particular  P. W\"olfle for useful discussions.
Also we thank A.~D. Mirlin for a
discussion of the ``condition of proportional couplings'' and
M. Wegewijs for providing us with a copy of Ref.\onlinecite{romeike02}. 
Finally, we express our gratitude to J. Reichert and H. Weber for
supplying us with the experimental data. 

\section{appendix}

We derive Eq. \ref{e2} from Eq. \ref{e3}. The effective single
particle Hamiltonian of the scattering problem is of the form
$\mathbf{H}=\mathbf{H_0} + \mathbf{H'}$, where 

\begin{equation}
  \mathbf{H_0} = \left( \begin{array}{ccc}
               h_L & 0 & 0 \\ 0 & h_C & 0 \\ 0 & 0 & h_R
	              \end{array} \right) \qquad
  \mathbf{H'} = \left( \begin{array}{ccc}
               0 & t_{\rm L}^\dagger & 0 \\ t_{\rm L} & 0  & t_{\rm R}
               \\ 0 & t^\dagger_{\rm R} & 0
	              \end{array} \right) 
  \end{equation}
The Hamiltonian of the uncoupled left (right) lead has been denoted
by $h_L$ ($h_R$), the Hamiltonian of the central unit (molecule) is $h_C$ and
the matrices $t_{\rm L,R}$ are the couplings of the left and right leads to the
central unit. If the molecule is uncoupled, the scattering states are
of the form
\begin{equation}
  \Phi_{ l}=(\phi_{ l},0,0)\qquad \Phi_{ r}=(0,0,\phi_{ r}).
\end{equation}
When the coupling has been switched on, the new
scattering states $\Psi_{ l,r}$ can be obtained from a
Lippmann-Schwinger equation
\begin{equation}
  \Psi_{ l,r} = \Phi_{ l,r} +
  \mbox{\boldmath ${\cal G}$\unboldmath}(E+i0) \ {\mathbf H'} \
  \Phi_{ l,r}
  \label{e21}
  \end{equation}
where $E$ labels the energy of the state and the resolvent matrix 
\boldmath${\cal G}$\unboldmath$(z)=[ z-{\mathbf H}]^{-1}$ has been
introduced. The lesser function is defined by (x=({\bf x},t))
\begin{equation}
  \mbox{\boldmath${\cal G^<}$\unboldmath}({\bf x},{\bf x'},E)
  = \sum_{\alpha=l,r} \int \! d(t{-}t') \ 
  \Psi_\alpha(x) \Psi^*_\alpha(x') \ f_\alpha \ 
  e^{\imath E(t-t')}.
    \end{equation}
Using Eq. \ref{e21} a matrix representation can be derived
\begin{equation}
  \mbox{\boldmath${\cal G^<}$\unboldmath}({\bf x},{\bf x'},E)=  
  \langle {\bf x} | [\mathbf{1} +
  \mbox{\boldmath${\cal G}$ \unboldmath} \mathbf{H'} ]
   \ \mathbf{g}^<_0 \ 
	  [\mathbf{1} + {\mathbf H'}\mbox{\boldmath${\cal G^\dagger}
  $\unboldmath}] | {\bf x'} \rangle
	  \label{e23}
    \end{equation}
with
\begin{equation}
    \mathbf{g}^<_0
               = \left( \begin{array}{ccc}
               g^<_L & 0 & 0 \\ 0 & 0 & 0 \\ 0 & 0 & g^<_R
	              \end{array} \right) 
\end{equation}
and 
\begin{equation}
  g^<_L({\bf x},{\bf x'},E)  = \sum_{l} \int \! d(t{-}t') \ 
  \phi_{ l}(x) \phi^*_{ l} (x') \ f_{ l} \ 
  e^{\imath E(t-t')}.
    \end{equation}
and a similar equation for $g^<_R$. In Eq. \ref{e1} only those
matrix elements of \boldmath${\cal G}^<$\unboldmath are needed,
that connect the left and right boundaries of the central unit. These elements
are given by the term in  Eq. \ref{e23} quadratic in the external
coupling. Since
\begin{equation}
  {\mathbf H'\ g^<_0 \ H'} = \left( \begin{array}{ccc}
                0 & 0 & 0 \\ 0 & \Sigma^<_C & 0 \\ 0 & 0 & 0
	              \end{array} \right) 
\end{equation}
with $\Sigma^<_{\rm C} = t_{\rm L} g^<_L t^\dagger_{\rm L} + t_{\rm R}
g^<_R t^\dagger_{\rm, R}$ and
since \boldmath ${\cal G}_0$ \unboldmath is diagonal one has
for the matrix element representing the central unit
\begin{equation}
  {\cal G}^< = {\cal G} \Sigma^<_{\rm C} {\cal G}^\dagger
  \end{equation}
and thus recovers Eq. (\ref{e2}). A similar derivation has
also been given by Brandbyge {\it et al.} \cite{brandbyge02}.

\end{multicols}
\end{document}